\documentclass[10pt,journal,twocolumn]{IEEEtran}
\usepackage{amsmath,amsfonts}
\usepackage{amsmath}
\usepackage{bm}
\usepackage{algorithmicx}
\usepackage{algorithm}
\usepackage{algpseudocode}
\usepackage{array}
\usepackage[caption=false,font=normalsize,labelfont=sf,textfont=sf]{subfig}
\usepackage{textcomp}
\usepackage{tcolorbox}
\usepackage{color}

\usepackage{stfloats}
\usepackage{url}
\usepackage{verbatim}
\usepackage{graphicx}
\usepackage{cite}
\usepackage{stfloats}
\usepackage{multirow}
\usepackage{makecell}
\begin{document}

\title{6DMA-Enabled ISAC for Low-Altitude Economy}

\author{Yingchao Jiao,~\IEEEmembership{Graduate Student Memebr,~IEEE},
        Xuhui Zhang,~\IEEEmembership{Memebr,~IEEE},\\
        Chunjie Wang,~\IEEEmembership{Graduate Student Memebr,~IEEE},
        Shuqiang Wang,~\IEEEmembership{Senior Memebr,~IEEE},\\
        Yanyan Shen,~\IEEEmembership{Memebr,~IEEE},
        Kejiang Ye,~\IEEEmembership{Senior Memebr,~IEEE}, and
        Chengzhong Xu,~\IEEEmembership{Fellow,~IEEE}
        % <-this % stops a space

\thanks{Yingchao Jiao and Chunjie Wang are with the Shenzhen Institutes of Advanced Technology, Chinese Academy of Sciences, Guangdong 518055, China, and also with the University of Chinese Academy of Sciences, Beijing 100049, China (e-mail: yc.jiao@siat.ac.cn; cj.wang@siat.ac.cn).}
\thanks{
Xuhui Zhang is with the Shenzhen Future Network of Intelligence Institute, the School of Science and Engineering, and the Guangdong Provincial Key Laboratory of Future Networks of Intelligence, The Chinese University of Hong Kong, Shenzhen, Guangdong 518172, China (e-mail: xu.hui.zhang@foxmail.com).
}

\thanks{
Shuqiang Wang, Yanyan Shen, and Kejiang Ye are with Shenzhen Institutes of Advanced Technology, Chinese Academy of Sciences, Guangdong 518055, China (e-mail: sq.wang@siat.ac.cn; yy.shen@siat.ac.cn; kj.ye@siat.ac.cn).
}
\thanks{Chengzhong Xu is with the State Key Laboratory of Internet of Things for Smart City, Department of Computer Science, University of Macau, Macau, SAR, China (e-mail: czxu@um.edu.mo).}
}

\maketitle

\begin{abstract}
In this paper, we investigate a six dimensional movable antenna (6DMA) enable integrated sensing and communications (ISAC) network in low-altitude economy.
The studied 6DMA can move in a three-dimensional space and rotate around its surface center, making full use of spatial freedom to adapt to the different location distributions of uncrewed aerial vehicles (UAVs) adjust channel conditions in time.
However, since the rotation and location change of 6DMA requires the assistance of a physical device, it is unreasonable for 6DMA to change locations too frequently.
Therefore, we propose a hierarchical deep reinforcement learning algorithm based on twin delayed deep deterministic policy gradient. The first layer optimizes the rotation and location of 6DMA with infrequent updates, and the second layer optimizes the UAV flight direction and base station transmit beamforming matrix in each time slot. Under the condition of satisfying the sensing intensity of the sensing target, the total data transmission rate to the UAVs is maximized.
The numerical results show that the proposed 6DMA-enable ISAC algorithm through joint optimization of multiple variables performs significantly better than the partially movable scheme and the fixed antenna position scheme.
\end{abstract}


\section{Introduction}

\IEEEPARstart{W}{ith the} {rise of the sixth-generation (6G) network and the rapid development of consumer-oriented uncrewed aerial vehicle (UAV) technology, an increasing number of UAVs have joined the low-altitude wireless network (LAWN), greatly expanding its applications in low altitude space \cite{10879807, liu2025movable}. The potential economic benefits of these applications are collectively referred to as the low-altitude economy (LAE). 
As a new growth point for the consumer electronics industry to expand into the three-dimensional (3D) space, the LAE has key applications such as aerial photography UAV, logistics distribution, and urban air mobility. Essentially, these are the evolutions of mobile consumer electronics terminals in low altitude scenarios.
For instance, UAVs enable real-time aerial monitoring, smart logistics, and emergency response, revolutionizing urban and rural LAE operations over LAWNs \cite{10606316}.
However, the successful implementation of the LAE requires the safe operation of various devices. Providing seamless high-quality wireless communication and ubiquitous sensing for low-altitude UAVs is of great importance \cite{11359737, wu2025low}. Consequently, there is a surge in the demand for connection and data exchange among devices, and traditional cellular networks may struggle to meet such demands \cite{9779853, liu2024energy, 11359731}.
}

{
To support LAE applications, integrated sensing and communication (ISAC) plays a pivotal role by providing ubiquitous sensing capabilities for UAVs while maintaining robust control and non-payload communication links \cite{10812728, 10693833}.
Specifically, the dual-functional radar-communication systems can significantly reduce the size, weight, and power constraints of aerial platforms, which is critical for UAV operations over LAWNs \cite{8972666}.
The key advancement of ISAC lies in resource sharing and joint design, reusing the spectrum, hardware, and signal-processing procedures. By leveraging communication signals in high frequency bands, it can simultaneously undertake sensing tasks, achieving synergy between communication and sensing \cite{9727202, 10233771}.
By enabling ISAC technology, the base station (BS) multiplexes communication echoes to exchange data with communication targets while completing the sensing task, providing an efficient solution \cite{9456851, 10224278}.
However, balancing and optimizing the two competing performances of communication and sensing has become a new problem \cite{10107972}. It is crucial to improve the communication rate as much as possible while maintaining a certain sensing intensity, especially in the face of the high dynamics of UAV-enabled wireless networks \cite{9916163, 11216153}.
}

{
To coordinate the communication and sensing performance, more precise adjustment of the antenna beam direction by the BSs is required \cite{11007274, 11374005}.
Traditional BSs equipped with fixed-position antennas (FPA) cannot flexibly allocate antenna resources to meet the demands at different times \cite{11154927}. Therefore, movable antennas (MAs), also known as fluid antennas, have been introduced into the wireless communication system \cite{10508218, 11247926}. There are linear movable antenna arrays, movable antenna planes, and three-dimensional movable antennas according to the different moving dimensions of MA antenna elements. Its core lies in dynamically adjusting the antenna array configuration by real-time sensing of the channel state, more accurately adjusting the beam direction, and achieving the collaborative optimization of communication and sensing beams \cite{11078433}. This not only ensures the communication rate but also improves the sensing accuracy, realizes the dynamic matching between multiple antennas and the channel environment, and provides a technical basis for scenarios such as autonomous driving, intelligent transportation, and secure communication \cite{10749968, 11374087, 11389918}.
}

\subsection{{Related Work}}
\subsubsection{{UAV-Enabled Systems over LAWNs}}
{
Due to their high flexibility and mobility, UAVs have been widely introduced in LAWNs. As aerial nodes, they establish reliable line-of-sight (LoS) wireless connections with BSs and ground users for data transmission, meeting the massive data-exchange demands in the LAE \cite{11373590,11122485}. 
Recent work has explored the optimization of various system performances in UAV-enabled LAWN.
In \cite{10949220}, the authors investigated a covert-environment backscatter communication scheme with a UAV relay as the radio-frequency source.
The scenario of random arrival and movement of tasks of multiple mobile users in a UAV-assisted mobile edge computing system was studied in \cite{10896833}. 
Regarding the uncertainties in LAWN caused by physical factors, \cite{11355698} achieved real-time recovery and optimization of beamforming performance under perturbations through the proposed algorithm with a long-short-term memory structure.
In \cite{11288100}, the authors proposed a multi-agent deep reinforcement learning (MADRL) scheme for the scenario where line of sight (LoS) and non-LoS co-exist, optimizing the resources of multiple UAVs and intelligent reflecting surfaces (IRSs) to maximize the system sum-rate.
In addition, aiming at the multi-target wireless sensing requirements of low-altitude wireless consumer networks, \cite{11328802} effectively improved the sensing accuracy and reliability through MAs.
In the UAV-enabled data collection scenario, the authors in  \cite{10980172} aimed to minimize the average age of information of ground devices and optimized the UAV's trajectory and collection scheduling.
}

{
However, most existing studies focus solely on optimizing single-system performance, neglecting ISAC systems, a critical enabler of LAE, and overlooking the essential need to jointly optimize communication and sensing performance for the next-geneartion LAWNs.
}

\subsubsection{{ISAC}}
{
ISAC has emerged as a landmark technology for 6G, aiming to utilize unified hardware and spectrum resources to perform both wireless communication and sensing \cite{9737357, wang2025joint}. The fundamental motivation behind ISAC is to address the spectrum congestion and to exploit the high-frequency bands that offer high resolution for sensing tasks \cite{10012421}.
The authors in \cite{11396947} proposed an IRS-assisted joint beamforming strategy and verified the effectiveness of this scheme in the communication-sensing coordination of LAWN.
\cite{10380513} optimized the transmit beamforming of multi-antenna BSs to maximize the minimum detection probability of the target area while ensuring communication quality.
\cite{11298206} took full-duplex autonomous aircraft as the low-altitude platform, it reduced the system energy consumption of the integrated communication-sensing-computing system in LAWN by optimizing resource allocation and task allocation.
In a system integrating sensing, communication, and power transmission, the performance trade-off among sensing, communication, and powering was achieved by setting the boundary of the Cramér-Rao bound (CRB) rate energy area \cite{10382465}.
To further optimize the performance of ISAC systems, research has pivoted towards sophisticated beamforming and resource allocation strategies. For instance, The authors in \cite{10217169} explored the fundamental trade-offs between sensing and communication metrics, such as the CRB for sensing and the achievable rate for communication, proposing joint waveform design methods to balance these competing objectives.
}

{
However, the high integration of ISAC technology also brings new problems. To meet the throughput requirements of ISAC for LAE, it is inevitable to deploy more antennas to serve highly mobile low-altitude devices. This incurs higher hardware costs and energy consumption. To address this challenge, it is necessary to seek high-performance and cost-effective multi-antenna technologies.
}

\subsubsection{{MA-Enabled Systems}}
{
Recently, the integration of ISAC with MAs has become a research hotspot \cite{liu2025uav, 11240557}. For instance, \cite{10955337} specifically discussed the opportunities of ISAC in 6G-enabled LAE, noting that high-mobility UAVs require more flexible beam-steering capabilities to maintain link reliability and sensing accuracy. 
Specifically, the authors of \cite{11177504} studied the cooperative trade-off of the MAs on sensing accuracy and communication quality. 
In the MA-enabled maximizing channel capacity, the authors in \cite{10870338} focused on optimizing the MA position and beamforming in clutter interference environment to maximize communication capacity and the authors in \cite{11178233} utilized an alternating optimization (AO) algorithm to jointly optimize the MA position, the UAV trajectory, and the beamforming to maximize user throughput.
The article \cite{10654366} aimed at the UAV-enabled wireless network enhanced by the MA array, maximize the user's total available data rate.
However, most MA techniques are limited to position adjustment in the plane, which is difficult to adapt to the requirements of non-uniform user distribution and complex sensing scenarios. Recently, some articles have proposed six-dimensional MAs (6DMA) to further break through the limitations of spatial freedom and provide ISAC with more flexible resource allocation capabilities. The authors in \cite{10945745} designed a 6DMA BS architecture, and the AO algorithm was designed to jointly optimize the three-dimensional position and rotation of the MA surface to improve the average user sum rate of the network.
The literature \cite{10918750} proposed a 6DMA enhanced anti-jamming scheme for cellular-connected UAV communication, using a block coordinate descent optimization algorithm designing the antenna position, the antenna rotation, and the beamforming to maximize the UAV reception signal-to-interference-and-noise ratio (SINR).
}

{
The introduction of the 6DMA mechanism increases the initial hardware cost and mechanical maintenance complexity on the BS, but the significant gains in system total rate and sensing intensity it brings mean that a single BS can serve more consumer terminals simultaneously.
Although replacing FPA with MA or 6DMA in LAWNs presents a promising opportunity to enhance performance, it introduces significant computational challenges due to the high-dimensional MA position optimization. }\par

\subsection{{Motivation and Contribution}}
{
With the rapid growth in the number of consumer-grade low-altitude devices and the emergence of an increasing number of LAE scenarios, efficient solutions are urgently needed to meet the high real-time requirements of LAWN. Although previous studies have explored the application of MA in real-time wireless networks \cite{11346858, 11316665}, most of them focus on linear antenna arrays, where the antennas can only adjust their two-dimensional positions, thus having limited utilization of spatial degrees of freedom. 
The 6DMA technology expands the spatial regulation dimensions, due to the increase in the dimensions and types of variables that need to be optimized, but the joint optimization of its six dimensional variables poses great challenges. 
Currently, most research adopts iterative optimization methods \cite{10918750, 11148174}, which show certain limitations in real-time scenarios and do not consider the impact of physical constraints of mechanical adjustments on performance.
}

{
Based on the above consideration, this paper explores the optimization of system performance in a 6DMA-enabled real-time ISAC system for LAE. We are mainly committed to jointly optimizing the trajectory of the UAV, the position and rotation of 6DMA, and the transmit beamforming of the base station, while ensuring the minimum sensing accuracy, so as to maximize the system's downlink data transmission rate. Therefore, a hierarchical deep reinforcement learning (DRL) algorithm leveraging twin delayed deep deterministic (TD3) is proposed.
The simulation results show that our proposed algorithm can play an effective role in the 6DMA-assisted low-altitude real-time network, and the proposed algorithm significantly outperforms other benchmarks.
The key contributions are summarized below:
}
\begin{itemize}
    \item {We introduce 6DMA into the LAWNs, where the BS equipped with 6DMA conducts downlink communication with UAVs and simultaneously uses the communication signal echoes to sense aerial targets. We formulate an optimization problem to jointly optimize the trajectories of UAVs, the position and rotation of 6DMA, and the transmit beamforming matrix of the BS, aiming to maximize the communication rate between the BS and UAVs while satisfying the sensing accuracy constraint.}
	\item {Since the update frequency of 6DMA is lower than that of UAVs trajectory and transmit beamforming matrix, we design a hierarchical DRL algorithm to optimize variables with different update frequencies in the optimization problem. In the first layer of the proposed algorithm, the TD3 algorithm is used to optimize the position and rotation of 6DMA, and in the second layer, the multi-agent TD3 (MATD3) algorithm is used to optimize UAV trajectories and transmit beamforming.}
	\item {We conduct extensive experiments during the training and validation processes. The numerical results show that, compared with benchmark methods, the proposed algorithm achieves a higher downlink communication rate. Moreover, we analyze the execution time of the algorithm and verify that the proposed algorithm can be applied to LAE scenarios with high real-time requirements.}
\end{itemize}

{
We organize the remainder of this article as follows.
Section \ref{sec:system model} introduces the 6DMA enabled ISAC scenario and describes the system model, communication model, sensing model, and then formulates the problem of maximizing the transmission rate. 
Section \ref{sec:proposed solution} transforms the problem into a Markov decision process (MDP), presents the proposed algorithm, and analyzes the computational complexity of the algorithm.
Section \ref{sec:numerical results} presents numerical results and Section \ref{sec:conclusion} concludes the paper.}

\section{System Model And Problem Formulation} \label{sec:system model}
\subsection{Network Model}
\begin{figure}[htbp]
    \centering
    \includegraphics[width=0.45\textwidth]{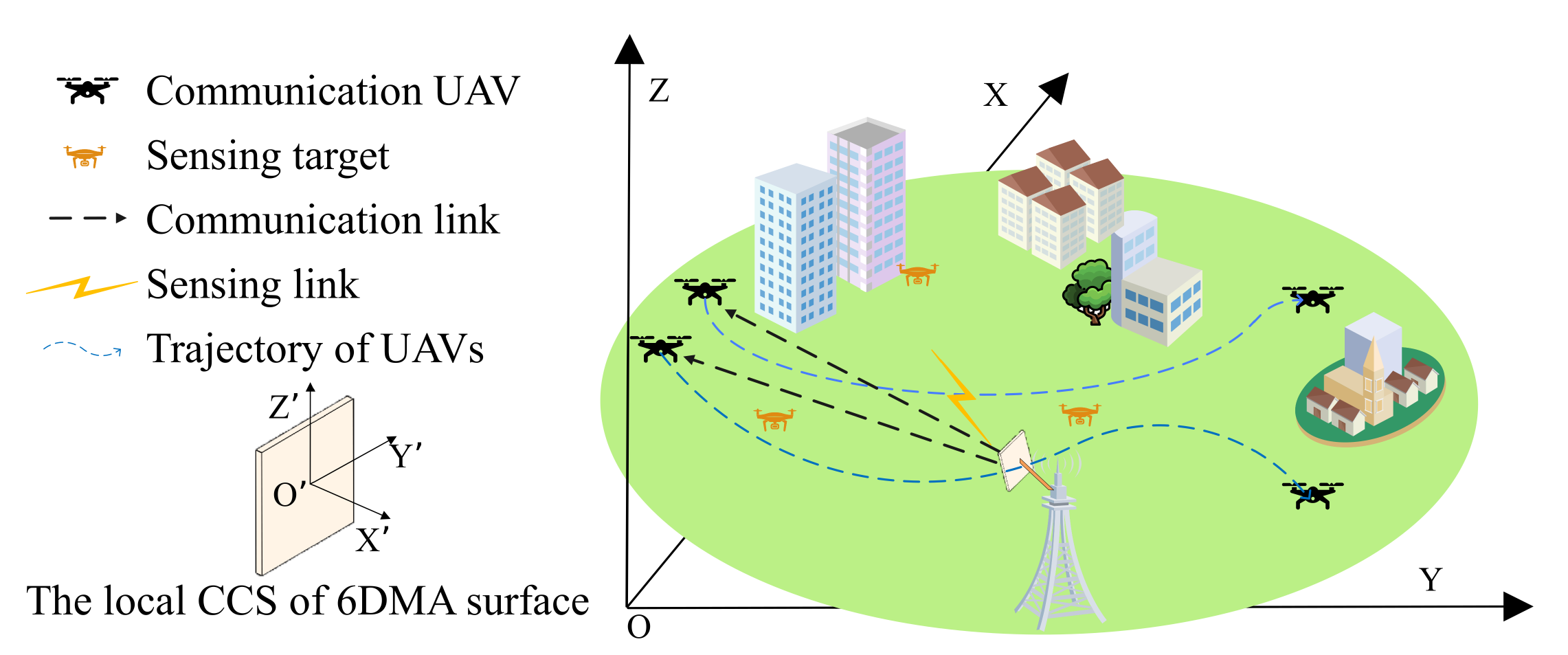}
    \caption{System model of the 6DMA-enabled ISAC.}
    \label{system_model}
\end{figure}

As shown in Fig.~\ref{system_model}, we consider a UAV-BS communication system. There is a BS equipped with a 6DMA surface. The 6DMA surface is assumed to be a UPA with $N$ MAs, the set of MAs is denoted as $\mathcal{N} = \{ 1, 2, \dots, N \}$. The antenna surface is connected to the BS via extendable and rotatable rods embedded with flexible wires, and thus its 3D positions and 3D rotations can be adjusted by the BS. The BS server $M$ UAVs with a single antenna for downlink communications and perceive $J$ sensing targets. 
The total flight time of all UAVs is $T$, divided into $K$ time slots, each with a length of $\Delta t = \frac{T}{K}$. The set of the time slots is denoted by a set $\mathcal{T} = \{ 1, 2, \dots, T\}$. 

\subsubsection{UAV trajectory model}
The UAVs are denoted by $m \in \mathcal{M} = \{1,2,\dots,M\}$. In the global cartesian coordinate system (CCS) $O-XYZ$, the position of the UAV is given by $p_m(t) = [x_m(t),y_m(t), z_m(t)]^{\mathsf{T}} $ in the time slot $t$, and the position of BS is denoted as $p_{BS} = \left[x_{BS}, y_{BS},0\right]^\mathsf{T}$. We keep $d_{\min}$ and $v_{\max}$ as the minimum distance between any two UAVs to avoid their collision and the maximum allowable velocity of the UAV $m$, respectively:
\begin{equation}
    \parallel p_{m_1}(t) - p_{m_2}(t) \parallel_2 \geq d_{\min},\ m_1,m_2 \in \mathcal{M},
    \label{min_distance}
\end{equation}
\begin{equation}
    \frac{\parallel p_m(t+1) - p_m(t) \parallel_2}{\Delta t} \leq v_{\max},\ \forall m \in \mathcal{M}.
\end{equation}

In particular, suppose that the BS simultaneously senses $J$ fixed position targets in the air, and the position of sensing targets denoted as $p_j = [ x_j, y_j, z_j]^\mathsf{T}, \forall j \in \mathcal{J}$, where $\mathcal{J} = \{1, 2, \dots, J\}$.

\subsubsection{6DMA model}
In particular, the position of the 6DMA surface center in the global CCS can be characterized by $p_a(t) = \left[ x_a(t), y_a(t), z_a(t)\right]^\mathsf{T} $. To adjust the position of 6DMA, BS adjusts the position $p_a(t)$ of 6DMA by rotating the rod every $T_r$ time slots. 
To characterize the rotation of the 6DMA array, we consider the local CCS $O'-X'Y'Z'$ with its origin being the center of the 6DMA array. The array rotation vector (ARV) is defined as $a(t) = \{\theta_x, \theta_y, \theta_z\}$ , where $\theta_x$, $\theta_y$ and $\theta_z$  denote the array rotation angles with respect to the $X-$axis, the $Y-$axis and the $Z-$axis at time slot $t$ , respectively, and the angle of rotation in the counterclockwise direction is positive. Then, the 6DMA array rotation matrices is given by
\begin{equation}
    \mathcal{R}(a(t)) = \mathcal{R}_{\theta_z}\mathcal{R}_{\theta_y}\mathcal{R}_{\theta_x}, 
\end{equation}
with
\begin{equation}
    \mathcal{R}_{\theta_x} = 
    \left[
    \begin{array}{ccc}
         1 & 0 & 0  \\
         0 & \cos\theta_x & \sin\theta_x \\
         0 & -\sin\theta_x & \cos\theta_x
    \end{array}
    \right],
\end{equation}

\begin{equation}
    \mathcal{R}_{\theta_y} = 
    \left[
    \begin{array}{ccc}
         \cos\theta_y & 0 & \sin\theta_y  \\
         0 & 1 & 0 \\
         -\sin\theta_y & 0 & \cos\theta_y
    \end{array}
    \right],
\end{equation}

\begin{equation}
    \mathcal{R}_{\theta_z} = 
    \left[
    \begin{array}{ccc}
         \cos\theta_z & -\sin\theta_z & 0  \\
         \sin\theta_z & \cos\theta_z & 0 \\
         0 & 0 & 1
    \end{array}
    \right].
\end{equation}
\par

We define the direction perpendicular to the 6DMA surface and pointing outward as the normal vector, which in the local CCS is denoted as $\bar{\mathbf{n}}_{ma}$. Therefore, its direction in the global CCS can be expressed as:
{\setlength{\abovedisplayskip}{7pt}
\setlength{\belowdisplayskip}{7pt}
\begin{equation}
    \mathbf{n_{\mathrm{ma}}} = \mathcal{R}(a(t)) \bar{\mathbf{n}}_{\mathrm{ma}}.
\end{equation}}

Since the rotation of 6DMA is controlled by a mechanical method, and to prevent signal blockage caused by the antenna plane facing the BS, it is necessary to impose constraints on the rotation range of 6DMA:
{\setlength{\abovedisplayskip}{7pt}
\setlength{\belowdisplayskip}{7pt}
\begin{equation}
    0 < \theta < \theta_{\max},\theta \in a_t,
\end{equation}}where $\theta_{max}$ represents the maximum allowable rotation angle.

The local CCS of the 6DMA surface is denoted by $o'-x'y'z'$, with the center serving as the origin $o'$. Let $\bar{p}_n = [\bar{x}_n,\bar{y}_n,\bar{z}_n]^{\mathsf{T}}$ denotes the position of $n$-th antenna of the 6DMA surface in its local CCS. Then, the antenna position vector of the $n$-th antenna of the 6DMA surface denoted by $p_n(t) \in \mathbb{R}^3$ in the global CCS, can be expressed as
{\setlength{\abovedisplayskip}{7pt}
\setlength{\belowdisplayskip}{7pt}
\begin{equation}
    p_n(t) = p_a(t) + \mathcal{R}(a(t))\bar{p}_n,\ n\in \mathcal{N}.
\end{equation}}

Due to the antennas in 6DMA being deployed on a two-dimensional plane, this plane naturally divides the space into two parts. To avoid signal blockage, the communication UAV and the sensing target should always be located in the same half-space as the front side of the 6DMA surface where the antenna is situated. Therefore, we impose the following constraint on the rotation of the 6DMA:
\begin{equation}
    \mathbf{n}_\mathrm{ma} p(t) \geq 0,
    \label{half_space}
\end{equation}
where $p(t) \in \mathcal{P}(t)$ and $\mathcal{P}(t) = \{p_1(t), p_2(t),\dots, p_{M+J}(t)\}$.

\subsection{Channel Model}
For the 6DMA BS-assisted communication system, the channel between the BS and UAV is not only related to the positions of the UAV, but also depends on the rotation of the 6DMA surface. In this subsection, we consider downlink transmission and model the connection from BS to the communication UAV and target.\par
For time slot $t$, we consider $s_m(t)$ denote the desired information signal by UAV $m$ with $\mathbb{E}\{ \mathbf{S}_M(t) {\mathbf{S}_M(t)}^\mathsf{H}\} = \mathbf{I}_M$, where $ \mathbf{S}_M(t) = \{ s_1(t), s_2(t), \dots, s_M(t)\} \in \mathbb{C}^{M \times 1}$, $\mathbf{w}_m(t) \in \mathbb{C}^{N \times 1}$ denote the transmit beamforming vector. 
Accordingly, the signal transmitted by the BS in the time slot $t$ is given by
\begin{equation}
    \mathbf{x}[t] = \sum_{m = 1}^{M} \mathbf{w}_m(t) s_m(t),\ \forall t \in \mathcal{T}.
\end{equation}

\subsubsection{Communication Channel Model}
Let $\theta_m(t)$ and $\phi_m(t)$ denote the elevation and azimuth angles of departure (AoD) from the antenna surface to the UAV $m$, respectively. The pointing vector corresponding to the direction $(\theta_m(t), \phi_m(t))$ is thus defined as
\begin{equation}
\begin{aligned}
    \mathbf{f}_m(t) = [
    & \cos(\theta_m(t))\cos(\phi_m(t)),\cos(\theta_m(t))\sin(\phi_m(t)), \\
    & \sin(\theta_m(t))
    ]^{\mathsf{T}}.
\end{aligned}
\end{equation}
So, the difference in signal path distance between the $n$-th antenna and the origin of the 6DMA surface $o'$ is ${{\mathbf{f}_m(t)}^\mathsf{T}p_n(t)}$,
Thus, the array response vector between the BS and the UAV $m$ can be expressed as
{
\begin{equation}
    \mathbf{g}_m(t) = \left[
    e^{\mathsf{j}\frac{2\pi}{\lambda}{\mathbf{f}_m(t)}^\mathsf{T}p_1(t)}, \dots, e^{\mathsf{j}\frac{2\pi}{\lambda}{\mathbf{f}_m(t)}^\mathsf{T}p_N(t)}
    \right],\ m \in \mathcal{M},
\end{equation}}where $\lambda$ denotes the carrier wavelength.
Due to the high altitude above the ground, the BS is more likely to have a LoS channel with UAV $m$. Thus, the channel vector between the BS and the UAV $m$ is expressed as 
\begin{equation}
    \mathbf{h}_m(t) = 
    \frac{\lambda}{4 \pi d_m(t)}e^{-\mathsf{j}\frac{2 \pi}{\lambda}d_m(t)}\mathbf{g}_m(t),
\end{equation}
where $d_m(t) = \parallel p_a(t) - p_m(t) \parallel_2$ denotes the distance from the 6DMA surface to the UAV $m$.
Then, the receiving signal of the UAV $m$ from BS is expressed as
\begin{equation}
    \mathbf{y}_m(t) = \mathbf{h}_m(t) \mathbf{x}[t] + \mathbf{n}_c,
\end{equation}
where $\mathbf{n}_c \sim \mathcal{CN}(0,\sigma_c^2)$ represents the additive white Gaussian noise (AWGN) at the UAV $m$ with average power $\sigma_c^2$.  Thus, the signal-to-interference-and-noise ratio (SINR) received from BS at UAV $m$ is given by
{
\begin{equation}
    \gamma_m(t) = \frac
    {\mid{\mathbf{w}_m(t)}^{\mathsf{H}}\mathbf{h}_m(t)\mid^2}
    {\sum_{i \in \mathcal{M},i \neq m}\mid{\mathbf{w}_i(t)}^{\mathsf{H}}\mathbf{h}_m(t)\mid^2 +
    % {\bm{h}_m^t}^H \bm{R}_s[t] \bm{h}_m^t + 
    \sigma^2},\ m \in \mathcal{M}.
    \label{SINR}
\end{equation} 
}
Accordingly, the sum-rate of all UAVs can be calculated by
\begin{equation}
    R_\mathrm{total}(t) = 
    \sum_{m = 1}^M\log_2(1 + \gamma_m(t)).
\end{equation}

\subsubsection{Sensing Channel Model}
Next, we model the signals of the sensing targets received by the BS.
Similar to the communication channel model, we let $\theta_j(t)$ and $\phi_j(t)$ denote the elevation and azimuth angles of arrival (AoA) from the target $j$ to the BS, respectively. The pointing vector corresponding to the direction $(\theta_j(t), \phi_j(t))$ is thus defined as
{\setlength{\abovedisplayskip}{7pt}
\setlength{\belowdisplayskip}{7pt}
\begin{equation}
    \begin{aligned}
        \mathbf{f}_j(t) = [
        & \cos(\theta_j(t))\cos(\phi_j(t)),\cos(\theta_j(t))\sin(\phi_j(t)),\\
        & \sin(\theta_j(t))
        ]^{\mathsf{T}}.
\end{aligned}\end{equation}}
So, the difference in the signal path between $n$-th antenna and the center of the 6DMA surface $o'$ is $\mathbf{f}_j(t)^\mathsf{T}p_n(t)$. 
Thus, the array response vector between the BS and the target $j$ can be expressed as
{\setlength{\abovedisplayskip}{7pt}
\setlength{\belowdisplayskip}{7pt}
\begin{equation}
\begin{aligned}
    \mathbf{g}_j(t) = \left[
    e^{\mathsf{j}\frac{2\pi}{\lambda}{\mathbf{f}_j(t)}^\mathsf{T}p_1(t)},  \dots, e^{\mathsf{j}\frac{2\pi}{\lambda}{\mathbf{f}_j(t)}^\mathsf{T}p_N(t)}
    \right],\ j \in \mathcal{J},
\end{aligned}\end{equation}}{where $\lambda$ denotes the carrier wavelength.} 
Similarly, the channel vector between the BS and the target $j$ is expressed as $\mathbf{h}^j_t$. Thus, the reflected signal received at BS is
{\setlength{\abovedisplayskip}{7pt}
\setlength{\belowdisplayskip}{7pt}
\begin{equation}
    \mathbf{y}_j(t) = \mathbf{h}_j(t) \mathbf{x}[t] + \mathbf{n}_s,
\end{equation}}where $\mathbf{n}_s \sim \mathcal{CN}(0,\sigma_s^2\mathbf{I}_N)$ represents the additive white Gaussian noise (AWGN) at the BS with average power $\sigma_s^2$. Furthermore, the signal-to-noise ratio (SNR) of the sensing target $j$ is given by
{\setlength{\abovedisplayskip}{7pt}
\setlength{\belowdisplayskip}{7pt}
\begin{equation}
\begin{split}
    SNR_j 
    &= \frac{\mathbb{E} \{ \mid \mathbf{h}_j(t) \mathbf{x}[t] \mid ^2 \} }{\sigma_s^2}\\
    &= \frac{\mathbf{h}_j(t) \left(\sum_{m = 1}^{M} \mathbf{w}_m(t) {\mathbf{w}_m(t)}^{\mathsf{H}} \right) {\mathbf{h}_j(t)}^{\mathsf{H}}}{\sigma_s^2}.
\end{split}
\end{equation}}

Thus, the transmit power constraint is given by
{\setlength{\abovedisplayskip}{7pt}
\setlength{\belowdisplayskip}{7pt}
\begin{equation}
    \sum_{m=1}^M {\mathrm{tr}}\left(\mathbf{w}_m(t){\mathbf{w}_m(t)}^{\mathsf{H}}\right) \leq P_{\max},\ \forall t \in \mathcal{T},
\end{equation}}where $P_{\max}$ denotes the maximum transmit power of BS.

\subsection{Problem Formulation}
Our objective is to maximize the sum-rate of all UAVs within $T$ time slots, while satisfying the sensing SNR requirement, flight missions, collision avoidance, and maximum transmission power constraints.The UAV's movement directions $\mathbf{a}_m(t), \forall m \in \mathcal{M}$, the communication beamforming matrix $\mathbf{w}_m(t),\forall m \in \mathcal{M}$, the ARV of 6DMA $a(t)$, and the position of the 6DMA surface center $p_a(t)$ in the $t$-th time slot are jointly optimized. Mathematically, this optimization problem is formulated as:
\begin{subequations}
\begin{align}
(P1) &\max_{\mathbf{a}_m(t), \mathbf{w}_m(t), a(t), p_a(t),}  \mathbb{E}\left[\sum_{t=1}^TR_{\mathrm{total}}(t)\right]\label{p1a} \\
\mathrm{s.t.:} 
 & \parallel p_{m_1}(t) - p_{m_2}(t) \parallel_2 \geq d_{\min},\ m_1,m_2 \in \mathcal{M}\label{p1b}, \\
 & \frac{\parallel p_m(t+1) - p_m(t) \parallel_2}{\Delta t} \leq v_{\max},\ \forall m \in \mathcal{M} \label{p1c}, \\
 & {\parallel p_{n_1}(t) - p_{n_2}(t) \parallel_2 \geq \frac{\lambda}{2},\ n_1,n_2 \in \mathcal{N}\label{p1d},} \\
 & 0< \theta < \theta_{\max},\ \theta \in a_t\label{p1e}, \\
 & \mathbb{E}\left[{\frac{1}{J}}\sum_{j=1}^{J}\mathrm{SNR}_j(t)\right]\geq\Gamma_{\mathrm{min}},\ \forall t \in \mathcal{T}\label{p1f}, \\
 & \sum_{m=1}^M \mathrm{tr}\left(\mathbf{w}_m(t){\mathbf{w}_m(t)}^{\mathsf{H}}\right)\leq P_{\max},\ \forall t \in \mathcal{T}\label{p1g},\\
 & \mathbf{n}_\mathrm{ma} p(t) \geq 0,\ \forall p(t) \in \{p_1(t), p_2(t),\dots, p_{M+J}(t)\} \label{p1h},
\end{align}
\label{P1}\end{subequations}
with constraint (\ref{p1b}) refers to the flight constraints of the UAVs and sensing targets, while constraint (\ref{p1c}) limits the velocity of UAVs to its maximum allowable velocity. Constraint (\ref{p1d}) ensures the minimum distance for any two antennas to avoid the coupling effects where $\lambda$ is the communication wavelength between UAV and BS, constraint (\ref{p1e}) limits the rotation range of the 6DMA surface, constraint (\ref{p1f}) ensures the minimum SNR for sensing where $\Gamma_{\mathrm{min}}$ is the minimum SNR that meets the requirements, constraint (\ref{p1g}) limits the maximum transmit power of the BS, and constraint (\ref{p1h}) ensures no signal blocking.

\begin{figure*}[ht!]
    \centering
    \includegraphics[width=0.95\linewidth]{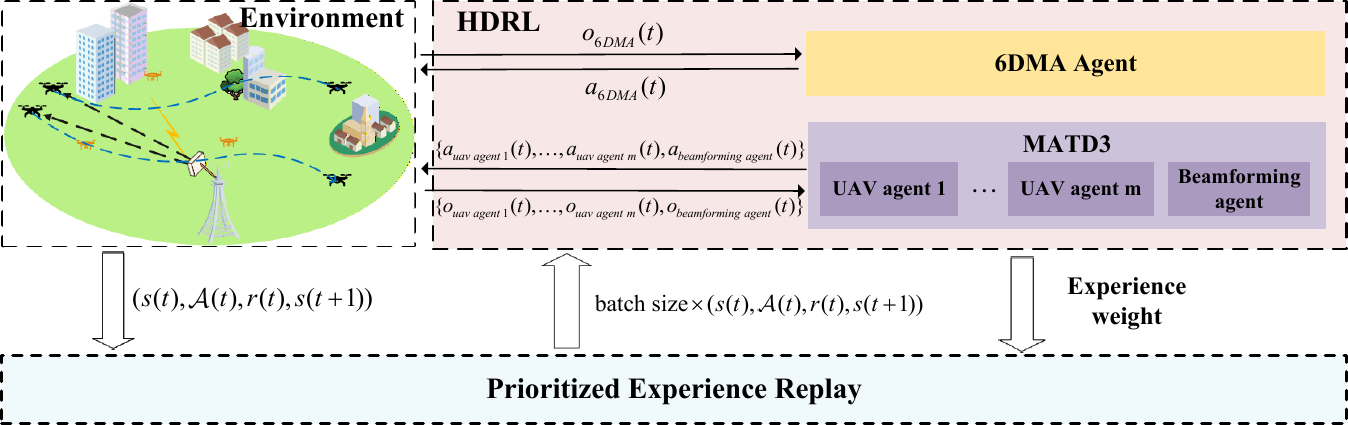}
    \caption{The structure of HDRL. In the training phase, a batch of experience sets is sampled from the prioritized experience replay to train agents, and the weight of this batch of experience is updated according to the loss value. In the execution phase, the HDRL framework obtains the current state from the environment, and each agent returns the corresponding action to the environment based on its own observation space.}
    \label{HDRL}
\end{figure*}

\section{Proposed Solution} \label{sec:proposed solution}
\subsection{Hierarchical DRL-based Algorithm}

In this section, we introduce the workflow of the proposed HDRL algorithm. In our proposed system model, since the 6DMA surface updates its state every $T_r$ time slots, and its reward is the average reward over the next $T_r$ time slots. However, UAV agents and beamforming agent need to give corresponding actions in each time slot to perform flight tasks and improve channel performance. Thus, there are two parts of HDRL: the first is single-agent DRL, which is used to update the rotation an position of the 6DMA surface every $T_r$ time slots, and the second part is MADRL, which is used to handle flight tasks and optimize the communication channel.\par
For the first part, the agent makes corresponding action according to the position information of UAVs in the current time slot, which represents the rotation angle and 3D location of the 6DMA surface in the next $T_r$ time slots. In order to accurately express the system performance of the action, the average of the UAV communication rate of the next $T_r$ time slots is used as a reward of the agent.
For the second part, a MADRL algorithm based on MATD3 is used. The algorithm adopts the network structure of centralized training with decentralized execution (CTDE). In order to avoid overestimation, each agent has two value networks, two target value networks, one policy network and one target policy network.\par
Due to the multidimensional nature of the optimization variables, different categories of agents are required to optimize different variables in MADRL. Based on the variables to be optimized in $(\ref{P1})$, the HDRL proposed in this article involves three types of distinct agents, each responsible for optimizing the position and rotation of the 6DMA, the flight speed and direction of the UAVs, and the beamforming of the antenna array, respectively.\par

\begin{table}
    \centering
    \renewcommand\arraystretch{1.25}
    \caption{MATD3 parameters}
    \label{parametemers of algorithm}
    \begin{tabular}{cc}
        \hline
        Parameters & Value\\
        \hline
        Number of training episode & 1000\\
        Learning rate of value network & 3e-4\\
        Learning rate of policy network & 1e-4\\
        Discount factor of reward $\gamma$ & 0.99\\
        Soft update factor of networks $\tau$ & 0.01\\
        Explore episode of agents & 600\\
        Action noisy standard deviation & 0.5\\
        Batch size of each training sample & 256\\
        \hline
    \end{tabular}
\end{table}

\subsection{Struct of MATD3}
\begin{figure*}[ht]
    \centering
    \includegraphics[width=0.95\linewidth]{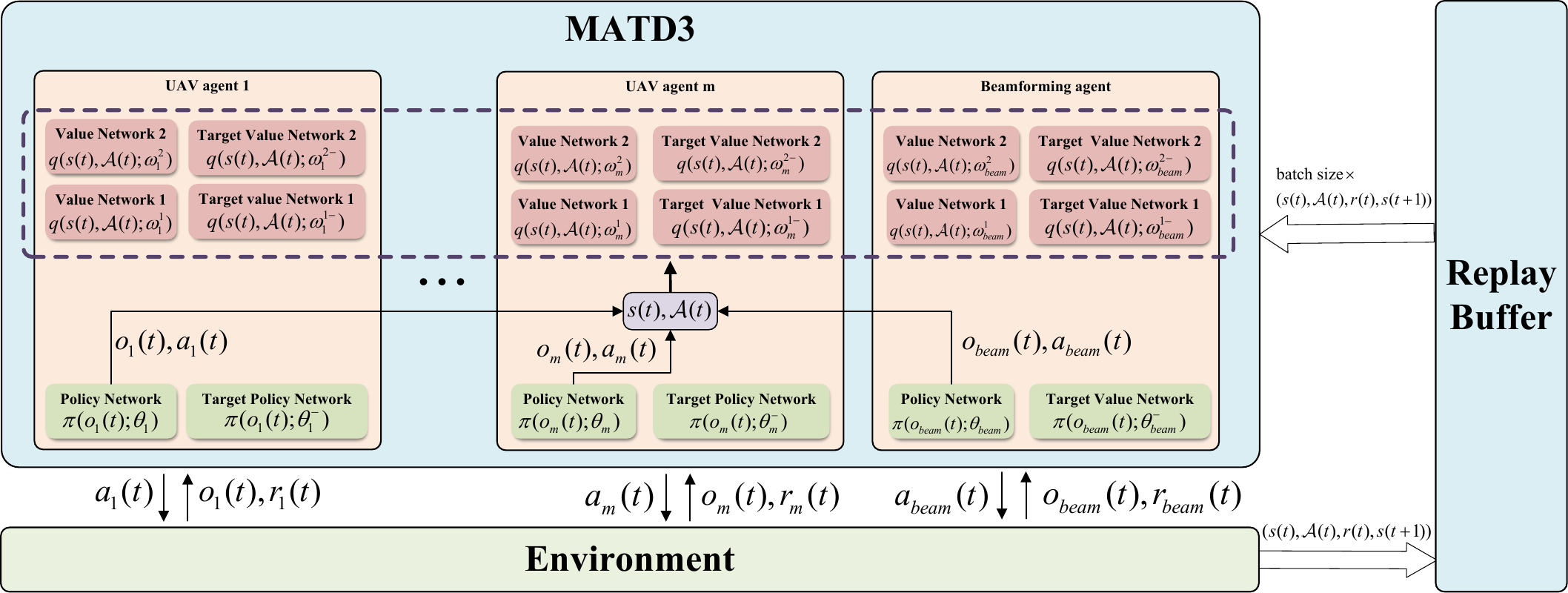}
    \caption{{The full workflow of MATD3. The MATD3 algorithm in this paper adopts the framework of CTDE. In the training stage, the observation and actions of all agents need to be integrated into $s(t)$ and $\mathcal{A}(t)$ as the input of the value networks to guide the networks update. In the execution stage, each agent only needs to use the policy network to give the corresponding action based on its own observation, and does not need value networks to participate in the decision-making process.}}
    \label{MATD3}
\end{figure*}
{
Due to the different update frequencies, the MADRL in this paper consists of $m$ UAV agents and one beamforming agent. Each agent uses the TD3 algorithm, with each agent corresponding to a policy network and two value networks: $\pi(o_i(t);\bm{\theta}_i)$, $q(s(t),\mathbf{a}(t); \bm{\omega}_i^1)$ and $q(s(t), \mathbf{a}(t);\bm{\omega}_i^2)$, and the corresponding target networks are  $\pi(o_i(t);\bm{\theta}_i^-)$, $q(s(t),\mathbf{a}(t);\bm{\omega}_i^{1-})$ and $q(s(t), \mathbf{a}(t);\bm{\omega}_i^{2-})$,
where $o_i$ denotes the observation of the $i$-th agent, $o(t) = \{o_1(t), o_2(t), \dots, o_M(t), o_{\mathrm{beam}}(t)\}$ is the set of observations for all agents, and $\mathbf{a}(t) = \{a_1(t), a_2(t), \dots, a_M(t), a_{\mathrm{beam}}(t)\}$ is the set of actions for all agents.
}\par
{
MADRL adopts the structure of CTDE, in which the policy network is deployed on each agent, the value network of all agents and the corresponding target networks are deployed on the central controller, and the central controller trains the value network with the observation and reward sent by the agent. The central controller feeds back the TD error to the agent for training the policy network. The specific structure of MATD3 is shown as Fig.~\ref{MATD3}.
}\par
{
During training, the target value network is used to make predictions and select the smaller value to calculate the TD target at agent $i$:
}
{
\begin{equation}
\begin{split}
    \hat{y}_i(t) = r_i(t) + \gamma \min(\hat{q}_{i}^{1-}(t+1),\hat{q}_{i}^{2-}(t+1)),&\\
    \forall i \in \{1, \dots, M+1\},&
\end{split}
\end{equation}}{where $r_i(t)$ is the reward value sample form experience replay buffer, $\hat{q}_{i}^{j-}(t+1) = q \left( s(t+1),\hat{\mathbf{a}}^{-}(t+1),\hat{\bm{\omega}}_{i,\text{now}}^{j-} \right), \forall j \in \{1,2\},\forall i \in \{1,\dots,M+1\}$, and $\hat{\mathbf{a}}_i^{-}(t+1)= \pi(o_i(t+1);\bm{\theta}_i)$. Thus, the TD error is calculated as
}
{
\begin{equation}
    \delta^{j}(t) = \hat{q}^j_{i}(t) - \hat{y}_i(t) , \forall j \in \{1,2\}.
\end{equation}
}
{
Then the parameters of the value networks are updated by minimizing the loss function as
}
{
\begin{equation}
\begin{split}
    \bm{\omega}_{i,\text {new}}^{j} \leftarrow \bm{\omega}^j_{i,\text {now}} - \alpha \delta^{j}(t) \bigtriangledown_{\bm{\omega}} q\left( s(t),a(t);\bm{\omega}^j_{i,\text {now}} \right),&\\
    \forall j \in \{1,2\},&  
\end{split}
\end{equation}
}
\par
In order to stabilize the output of the policy network, the policy is updated every $d$ steps, and is updated by
{
\begin{equation}
    \boldsymbol{\theta}_{i,\text {new}} \leftarrow \boldsymbol{\theta}_{i, \text{now}}+\beta \cdot \nabla_{\boldsymbol{\theta}} \boldsymbol{\mu}\left(s(t) ; \boldsymbol{\theta}_{i,\text {now}}\right) \cdot \nabla_{\boldsymbol{a}} q\left(s(t), \widehat{\boldsymbol{a}}(t) ; \bm{\omega}_{i,\text{now}}^1\right).
\end{equation}
}

{
Then the parameters of the target networks are updated by soft update as follows
}
{
\begin{subequations}
    \setlength{\abovedisplayskip}{2pt}
    \setlength{\belowdisplayskip}{2pt}
    \begin{align}
         \bm{\theta}^{-}_{i, \text{new}} &\leftarrow \tau \cdot \bm{\theta}_{i,\text{new}} + (1 - \tau) \cdot \bm{\theta}^{-}_{i, \text{now}}, \\
         \bm{\omega}^{1-}_{i, \text{new}} &\leftarrow \tau \cdot \bm{\omega}^1_{i, \text{new}} + (1 - \tau) \cdot \bm{\omega}^{1-}_{i, \text{now}}, \\
         \bm{\omega}^{2-}_{i, \text{new}} &\leftarrow \tau \cdot \bm{\omega}^2_{i, \text{new}} + (1 - \tau) \cdot \bm{\omega}^{2-}_{i, \text{now}}.
    \end{align}
\end{subequations}
}

\subsection{Problem Transformation}
Under normal circumstances, problem (\ref{P1}) can be converted into an MDP for solving. The framework employs different agents to optimize different variables, with the beamforming agent responsible for the transmission beamforming of BS, the 6DMA agent responsible for optimizing the movement strategy of 6DMA, and the UAV agents responsible for trajectory optimization of UAVs. We define basic MDP elements, including state, action and reward function as follows: \par
$\textbf{State\ Space}\ \mathcal{S}$:
In the $t$-th time slot, the observation of the UAV agent $m$ consists of the current time slot, the position of the center of 6DMA surface, the current position and end position of UAV $m$. And the observation of the beamforming agent is composed of the new channel state information (CSI) $H_{t} = \{ H_c(t), H_s(t)\}$. 
The observation of the 6DMA agent consists of the positions of BS and all UAVs. 
The observations of different agents are shown as
\begin{subequations}
    \setlength{\abovedisplayskip}{2pt}
    \setlength{\belowdisplayskip}{2pt}
    \begin{align}
        &o_m(t) = \{p_m(t), p_\mathrm{BS}, p_m^\mathrm{end}, t\}, m \in \mathcal{M},\\
        &o_\mathrm{beam}(t) = \{H_c(t), H_s(t)\},\\
        &{o_\mathrm{6DMA}(t) = \{p_\mathrm{BS}, p_1(t), p_2(t), \dots, p_M(t)\}}.
    \end{align}
\end{subequations}
\par
$\textbf{Action\ Space}\ \mathcal{A}$: 
In the $t$-th time slot, the action space $\mathcal{A}$ is composed of all possible actions that could be potentially executed by the agents. We define the movement directions $\bm{a}_m(t)$ and the velocity $v_m(t)$ as the actions of UAV agent $m$, the action of the beamforming agent defined as the communication beamforming matrix $\mathbf{w}_m(t)$. 
And the position of the center of the 6DMA surface $p_a(t)$ and ARV $a(t)$ as the actions of the 6DMA agent. 
Thus, the action of the agents at the $t$-th time slot can be denoted as
\begin{equation}
    \mathcal{A}(t) = \{ \mathcal{A}_m(t), \mathcal{A}_\mathrm{beam}(t), \mathcal{A}_\mathrm{6DMA}(t)\}, \forall m \in \mathcal{M},
\end{equation}
where $\mathcal{A}_m(t)=\{\bm{a}_m(t), v_m(t)\}$, $\mathcal{A}_\mathrm{beam}(t)=\mathbf{w}_m(t)$, $\mathcal{A}_\mathrm{6DMA}(t)=\{a(t),p_a(t)\}, t\%T_r = 0$.
\par
$\textbf{Reward\ Function}\ \mathcal{R}$: 
The reward function consists of a reward value and a penalty value. The reward value is the system transmission rate of the current time slot, and the penalty value is the punishment when agents fail to meet constraints.

\begin{itemize}
    \item First, let the optimization goal $R_\mathrm{total}(t)$ as the basic reward part, and the reward value of the reward function is given by $R_\mathrm{total}(t)$.
    \item Second, when the agents fail to meet the constraints of problem (\ref{P1}), the corresponding punishment is required. 
    When the constraint (\ref{p1b}) is not satisfied between any two UAVs, the corresponding punishment is $-\delta_1$, where $\delta_1 > 0$. The lager $\delta_1$ is, the more the agent attaches importance to the given constraint.
    And when the rotation of the 6DAM surface violates the constraint (\ref{p1h}) and causes signal blocking, the corresponding punishment is $-\delta_2$, where $\delta_2 > 0$.
    According to the constraint (\ref{p1f}), the agent needs to satisfy the preset sensing SNR requirement $\Gamma_{\min}$. In the time slot $t$, if the average sensing SNR of $J$ sensing targets is less than $\Gamma_{\min}$, the corresponding punishment is $\delta_3 = \max\{0, \Gamma_{\min} - \mathbb{E}[\frac{1}{J}\sum_{j = 1}^J SNR_j(t)]\}$, where $\delta_3 \geq 0$. 
    When the transmit power of the BS exceeds the maximum allowed transmit power (\ref{p1g}), the corresponding penalty term is $\delta_4 = \max\{0, \mathrm{tr}\left(\mathbf{w}_m(t) {\mathbf{w}_m(t)}^{\mathsf{H}}\right) - P_{\max}\}$, where $\delta_4 \geq 0$.
    {To encourage more precise adjustment of the pointing direction of the 6DMA surface, we set $\delta_5=\frac{1}{M}\cos\theta_t$, where $\theta_t$ is the average value of the angles between the normal vector of the 6DMA surface and the UAVs at time slot $t$.}
\end{itemize}

In particular, when the agents fail to meet the distance limit (\ref{p1b}) between UAVs and the distance limit (\ref{p1d}) between antennas, the system cannot complete communication and sensing tasks. Thus, the reward function of different agents is given by:
\begin{subequations}
    \setlength{\abovedisplayskip}{2pt}
    \setlength{\belowdisplayskip}{2pt}
    \begin{align}
        &r_m(t) = (1-\epsilon_1)R_\mathrm{total}(t) - \epsilon_1 \delta_1,\\
        &r_\mathrm{beam}(t) = R_\mathrm{total}(t) - \delta_3 - \delta_4 + \frac{1}{J} \sum_{j=1}^J SNR_j(t),\\
        &r_\mathrm{6DMA}(t) = (1-\epsilon_2)\sum_{t=i\times T_r}^{(i+1)\times T_r}R_\mathrm{total}(t) - \epsilon_2\delta_2 + \delta_5,
    \end{align}\label{reward_functions}\end{subequations}where $\epsilon_1$ is a binary variable, $\epsilon_1 = 1$ means that the actions not meet the constraint (\ref{p1b}), when $\epsilon_1 = 0$, it is the opposite. And the same applies to $\epsilon_2$ when it does not meet the constraint (\ref{p1h}).
And the main steps of HDRL algorithm can be summarized in Algorithm~\ref{algorithm1}.

\begin{algorithm} \scriptsize
	\caption{The HDRL Algorithm for Solving (P1)}
    \label{algorithm1}
		\begin{algorithmic}[1]
				\State Initialize the actor and critic networks, target actor and target critic networks;
				\State Initialize the replaybuffer $\mathcal{D}$;
				\For{episoed $e= 1,\dots,E$}
					\State Initalize environment;
					\For{time slots $t = 1,\dots,T$}
						\If{$t \% T_r == 0$}
							\State Get observation $o_\mathrm{6DMA}(t)$;
							\State 6DMA agent select action $\mathcal{A}_{6DMA}(t)$;
                            \State \parbox[t]{\dimexpr 0.87\linewidth-\algorithmicindent}{6DMA agent execute action $\mathcal{A}_\mathrm{6DMA}(t)$, receive reward $r_{6DMA}(t)$ at $t'=t+T_r$ and next observation $o_\mathrm{6DMA}(t')$;}
							\State \parbox[t]{\dimexpr 0.87\linewidth-\algorithmicindent}{Store transition ($o_\mathrm{6DMA}(t)$,$\mathcal{A}_\mathrm{6DMA}(t)$,\ $r_\mathrm{6DMA}(t)$,$o_\mathrm{6DMA}(t')$) in $\mathcal{D}$;}
							\State \parbox[t]{\dimexpr 0.87\linewidth-\algorithmicindent}{Sample a random batch transitions ($o_\mathrm{6DMA}(t),\ \mathcal{A}_\mathrm{6DMA}(t),r_\mathrm{6DMA}(t),o_\mathrm{6DMA}(t')$) from $\mathcal{D}$;}
							\State Update critic networks of 6DMA agent;
							\If{$(t \% T_r) \% d==0$}
								\State Update actor network of 6DMA agent;
								\State \parbox[t]{\dimexpr 0.8\linewidth-\algorithmicindent}{Update target networks of 6DMA agent by soft update;}
							\EndIf
						\EndIf
						\State \parbox[t]{\dimexpr 0.87\linewidth-\algorithmicindent}{Get observations $\mathcal{O}(t) = \{o_1(t),o_2(t),\dots,o_M(t),o_\mathrm{beam}(t)\}$;}
						\State \parbox[t]{\dimexpr 0.87\linewidth-\algorithmicindent}{MATD3 select actions $\mathcal{A}(t) = \{\mathcal{A}_m(t), \mathcal{A}_\mathrm{beam}(t), \forall m \in \mathcal{M}\}$;}
						\State \parbox[t]{\dimexpr 0.87\linewidth-\algorithmicindent}{MATD3 execute actions, received reward $\mathcal{R}_t = \{r_1(t),\dots,r_M(t),r_\mathrm{beam}(t),r_\mathrm{6DMA}(t)\}$ and next observations $\mathcal{O}(t+1)$;}
						\State Store transition ($\mathcal{O}(t),\mathcal{A}(t),\mathcal{R}(t),\mathcal{O}(t+1)$) in $\mathcal{D}$;
						\State \parbox[t]{\dimexpr 0.87\linewidth-\algorithmicindent}{Sample a random batch of $B$ transitions ($\mathcal{O}(t),\mathcal{A}(t),\mathcal{R}(t),\mathcal{O}(t+1)$) from $\mathcal{D}$;}
						\State \parbox[t]{\dimexpr 0.87\linewidth-\algorithmicindent}{Update critic networks of uav agents and beamforming agent;}
						\If{$e \% d == 0$}
							\State \parbox[t]{\dimexpr 0.87\linewidth-\algorithmicindent}{Update actor networks of uav agents and beamforming agent;}
							\State \parbox[t]{\dimexpr 0.87\linewidth-\algorithmicindent}{Update target networks of uav agents and beamforming agent by soft update;}
						\EndIf
					\EndFor
				\EndFor
	   \end{algorithmic}
\end{algorithm}

\subsection{Computational Complexity Analysis}
{
We let $O_\mathrm{dim}=\{o_{m},o_\mathrm{beam},o_\mathrm{6DMA}\}$, $\mathcal{A}_\mathrm{dim}=\{a_m,a_\mathrm{beam},a_\mathrm{6DMA}\}$, $C=\{ C_m,C_\mathrm{beam}, C_\mathrm{6DMA}\}$ represent the input dimensions, output dimensions and computational complexity of UAV agent $m$, beamforming agent and 6DMA agent, respectively. Let $g_1,g_2$ represent the dimensions of the hidden layers. 
The complexities of a single forward propagation of the policy network and the value network are $ C_\mathrm{policy} = \mathcal{O}(og_1 + g_1g_2 + g_2a )$ and $C_\mathrm{value} = \mathcal{O}(I_\mathrm{value}g_1 + g_1g_2 + g_2)$, respectively, where $I_\mathrm{value}$ = $m \times (o_m + a_m) + o_\mathrm{beam} + a_\mathrm{beam}$ denotes the input dimensions of the value networks of MATD3, and $I_\mathrm{value}$ = $o_\mathrm{6DMA} + a_\mathrm{6DMA}$ is the input dimensions of the value network of the 6DMA agent.}\par
{
As shown in Fig.~\ref{MATD3}, in the proposed scheme, each agent has one policy network, one target policy network, two value networks and two target value networks. 
Therefore, in each update round, the computational complexity of the value network update is $ C_\mathrm{value\:update} = \mathcal{O}(C_\mathrm{policy} + 8 \times C_\mathrm{value})$. Due to the delayed update strategy of the policy network, in each update round, the computational complexity of the policy network update is $C_\mathrm{policy\:update} = \mathcal{O}(\frac{1}{2} ( C_\mathrm{policy} + 3 \times C_\mathrm{policy} ) )$.
The computational complexity of each agent in each update round is $C = \mathcal{O}( C_\mathrm{value\:update} + C_\mathrm{policy\:update})$.}\par

Thus, the computational complexity of 6DMA agent is $\mathcal{O} ( E T B C_\mathrm{6DMA})$, and the total computational complexity of MATD3 is $\mathcal{O}( ETB ( m ( C_m) + C_\mathrm{beam} ) )$, where $E$ is the number of training rounds, $T$ is the number of time slots in each round, and $B$ is the batch size of each training sample. 
As a result, the computational complexity of HDRL is $\mathcal{O} (ETB (C_\mathrm{6DMA} + m ( C_m) + C_\mathrm{beam} ))$.

\section{Numerical Results} \label{sec:numerical results}
This section presents numerical results to validate the performance of proposed algorithm. 
In the simulation, consider a area of $500m \times 500m$ with $M = 4$ UAVs and $J = 3$ sensing targets, where UAVs fly in this area. The total flight time of the UAV is $ T = 300s$, and is divided into $K = 60$ time slots. There are $N = 4$ movable antennas initially uniformly distributed at the 6DMA surface, the side length of surface is $L = 1m$, and the initial position of the center of the 6DMA surface $o'$ in the global 3D CCS is $p^a_0 = [0, 0, 200]$. Table \ref{parametemers_of_SystemModel} summarizes the detailed parameters of the system model. 
\begin{table}
    \centering
    \renewcommand\arraystretch{1.25}
    \caption{System model parameters}
    \label{parametemers_of_SystemModel}
    \begin{tabular}{cc}
        \hline
        Parameters & Value \\
        \hline
        Carrier wavelength $\lambda$ & 0.125m\\
        Communication AWGN $\sigma_c$ & -50dBm\\
        Sensing AWGN $\sigma_s$  & -50dBm\\
        6DMA update frequency $T_r$ & 10\\
        Maximum speed of UAV $v_{\max}$ & 8m/s\\
        Minimum distance between any UAVs $d_{\min}$ & 3m\\
        Minimum SNR of sensing targets $\Gamma_{\min}$ & 1dB\\
        Maximum transmit power of BS $P_{\max}$ & 40mW\\
        Maximum rotation range per time $\theta_{\max}$ & $10^{o}$\\
        \hline
    \end{tabular}
    \vspace{-0.2cm}
\end{table}
Furthermore, the proposed scheme and the benchmark schemes are as follows. \\
(1) $\textbf{Scheme\ 1\ (Proposed)}$: Joint optimization of the UAV trajectory, the beamforming matrix, the rotation and position of the 6DMA surface. \\
(2) $\textbf{Scheme\ 2\ (TD3)}$: Each agent in the second layer of HDRL uses only its own observation during the training and excution processes. \\
(3) $\textbf{Scheme\ 3\ (6DMA\ with\ rotation\ only)}$: Fix the position of the center of the 6DMA surface $o'$, optimizing the rotation of the 6DMA surface only. \\
(4) $\textbf{Scheme\ 4\ (6DMA\ with\ circular\ movement\ only)}$: Fix the rotation of the 6DMA, optimizing the position with circular movement only. \\
(5) $\textbf{Scheme\ 5\ (FPA)}$: The rotation and position of the 6DMA surface remain unchanged, optimizing the trajectory of UAV and the beamforming matrix only.\par
{
The simulations in this work were conducted using NVIDIA GeForce RTX 5060Ti GPU and AMD 9700X 8-core CPU. The GPU is rquipped with 16 GB of memory, which provides sufficient capacity for thr computational demands of the proposed scheme. The driver version of GPU used is NVIDIA-SMI 591.74 and CUDA version 13.1. The data in Fig. 8-Fig. 10 and Table~\ref{inference_latency} for each scheme and the corresponding system parameters are the average of the data obtained from 20 executions of the model.}

\begin{figure}[htbp]
    \vspace{-0.2cm}
    \centering
    \includegraphics[width=0.8\linewidth]{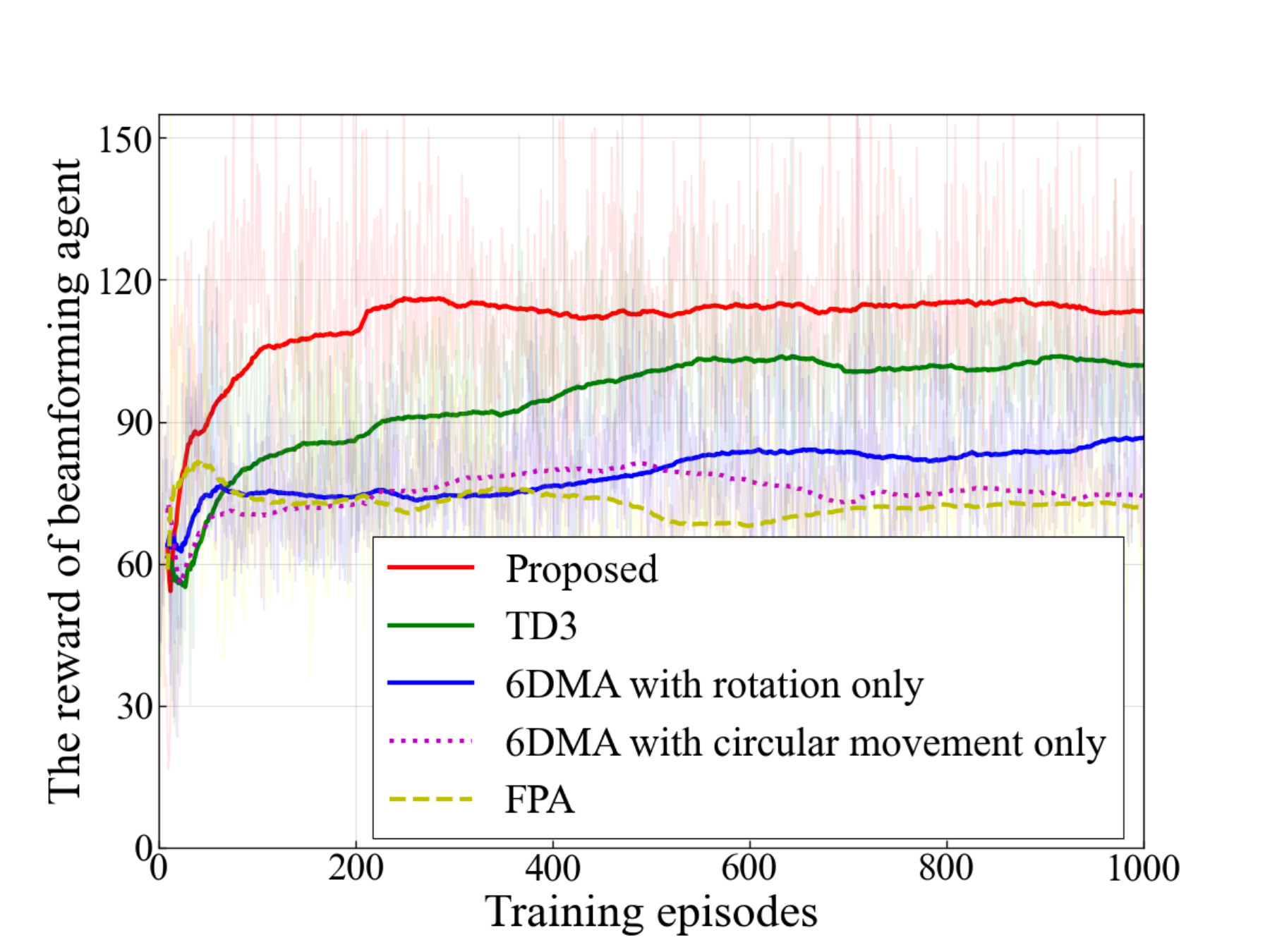}
    \caption{The reward value of beamforming agent versus the training episodes.}
    \label{reward_beam}
    \vspace{-0.2cm}
\end{figure}

\begin{figure}[htbp]
    \vspace{-0.2cm}
    \centering
    \includegraphics[width=0.8\linewidth]{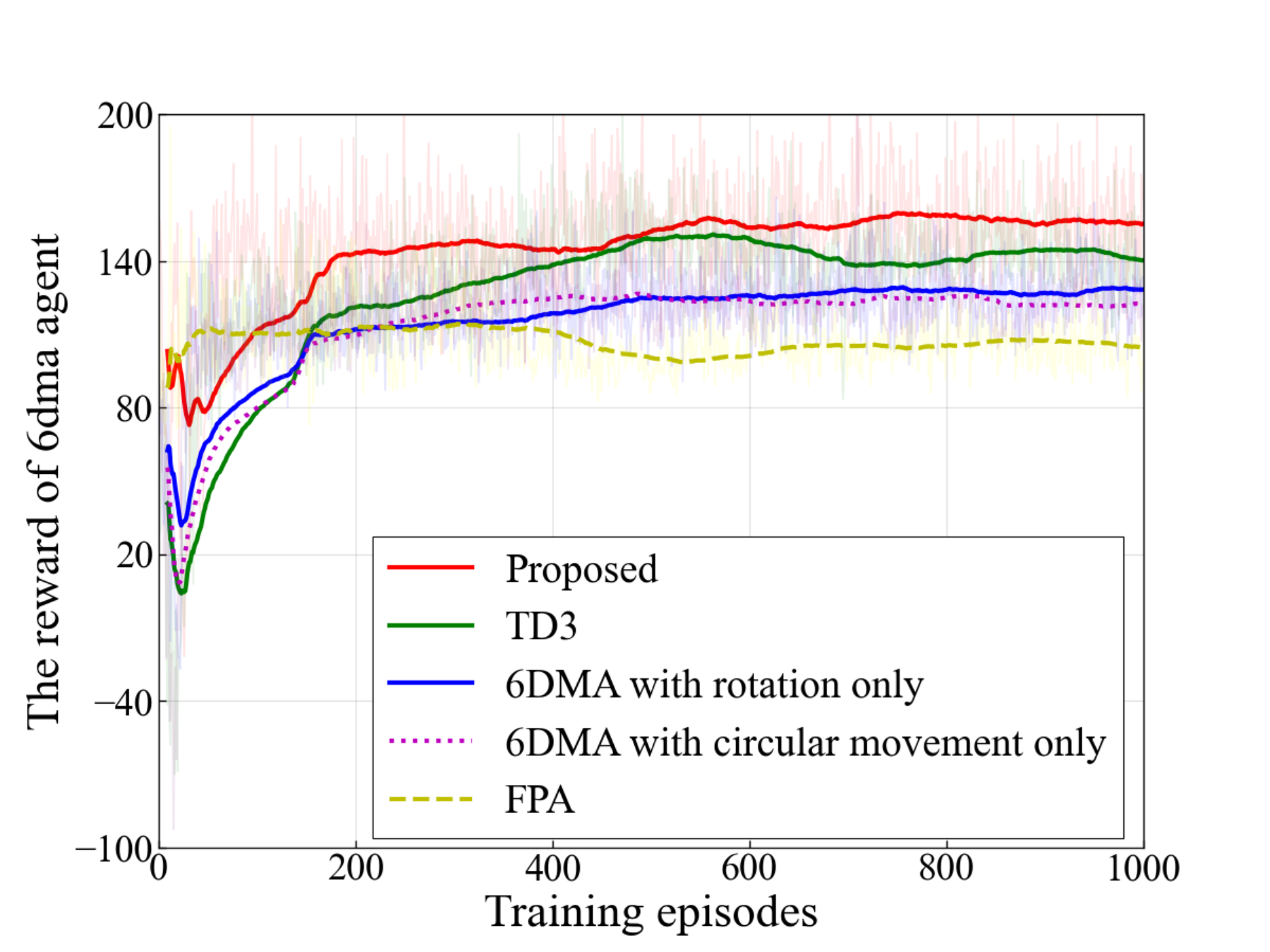}
    \caption{The reward value of 6DMA agent verus the training episode.}
    \label{reward_6dma}
    \vspace{-0.2cm}
\end{figure}

Fig. \ref{reward_beam} and Fig. \ref{reward_6dma} compare the convergence performance of the beamforming agent and 6DMA agent in the proposed algorithm and the comparative algorithms. The horizontal axis represents the training episodes, and the vertical axis represents the reward of the current agent. As a direct factor guiding agent training, it can be seen from the figures that each agent in the proposed algorithm has a faster convergence speed and higher training stability. This advantage stems from the fact that MATD3 can use the observations of all agents during the training process, resulting in more effective exploration and policy learning.
\begin{figure}[htbp]
    \vspace{-0.2cm}
    \centering
    \includegraphics[width=0.8\linewidth]{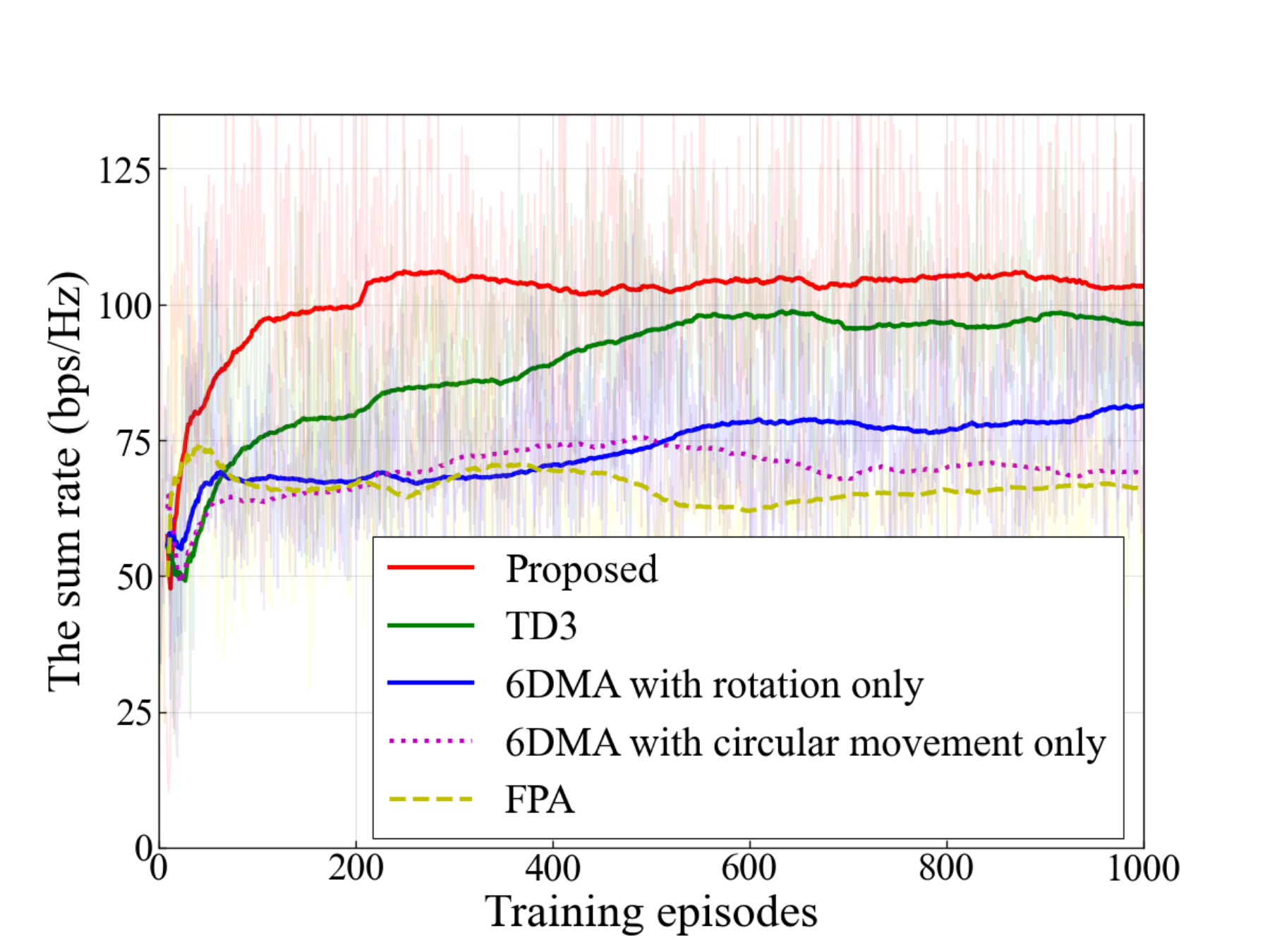}
    \caption{The sum rate verus the training episodes.}
    \label{sum_rate_convergence}
    \vspace{-0.2cm}
\end{figure}
\begin{figure}[htbp]
    \vspace{-0.2cm}
    \centering
    \includegraphics[width=0.8\linewidth]{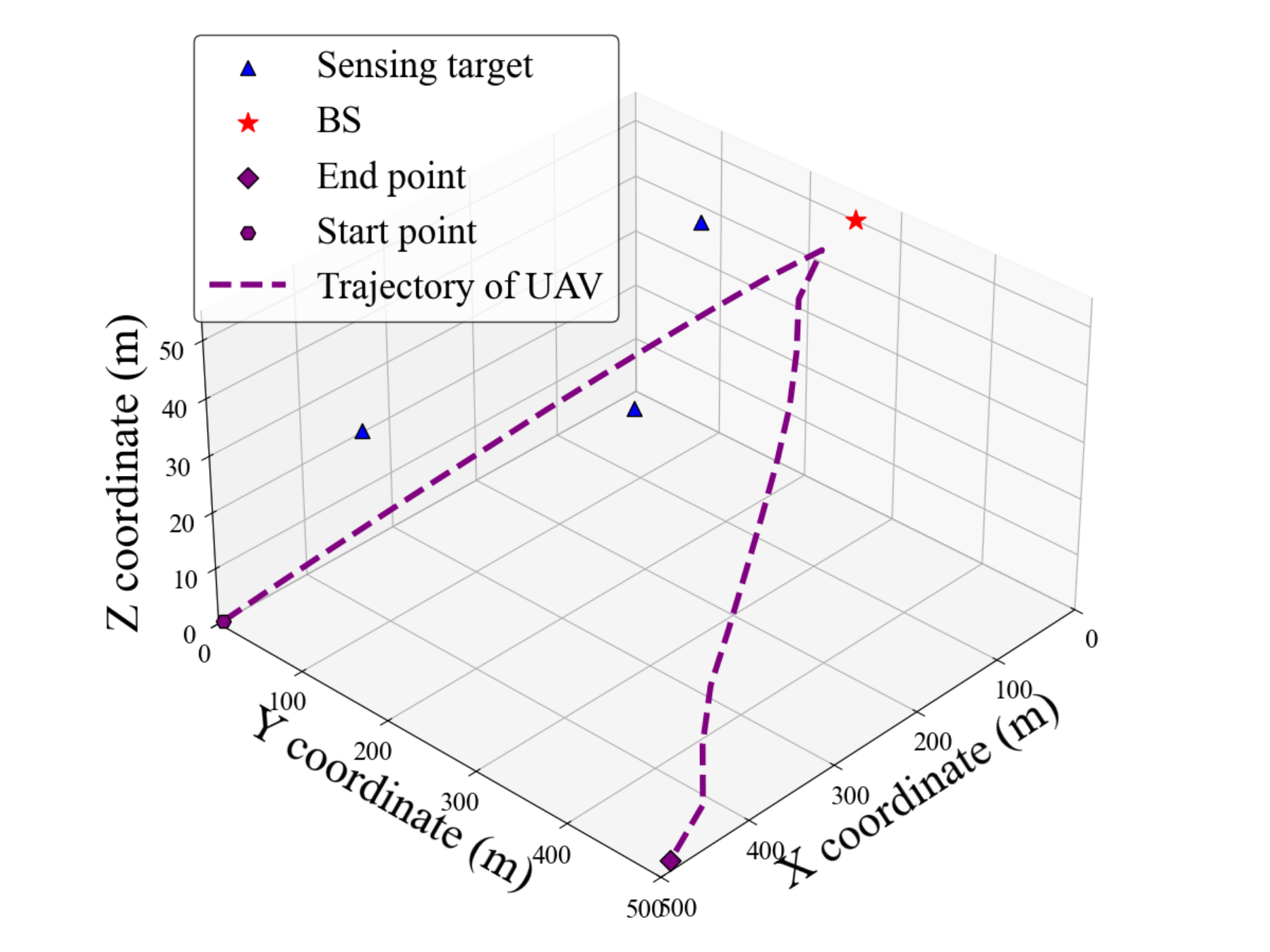}
    \caption{The trajectory of the first UAV.}
    \label{trajectory}
    \vspace{-0.2cm}
\end{figure}

And we evaluate the convergence of the proposed HDRL algorithm to optimize the objectives under joint optimization of different types of variables. As shown in Fig.~\ref{sum_rate_convergence}, it can be found that the algorithms under different schemes eventually converge to a stable range. And as can be seen from Fig.~\ref{sum_rate_convergence}, the performance of the proposed scheme is obviously better than other schemes, because the proposed scheme simultaneously optimizes the position and rotation of 6DMA, so that the transmitted signal of the BS can be more accurately pointed to the UAV group.
In order to reflect the performance of the proposed algorithm in other aspects, we show the three-dimensional flight trajectory of one of the UAVs in Fig.~\ref{trajectory}. It can be seen from the trajectory that the UAV starts from the initial point. During flight, the algorithm guides the UAV to fly close to the BS, and the UAV can reach the end point within the specified time. The results show that the proposed algorithm can optimize the UAV trajectory as much as possible and improve the effectiveness of the communication performance of the system on the basis of guiding the UAV to complete the original flight task.
\par

\begin{figure}[htbp]
    \centering
    \includegraphics[width=0.8\linewidth]{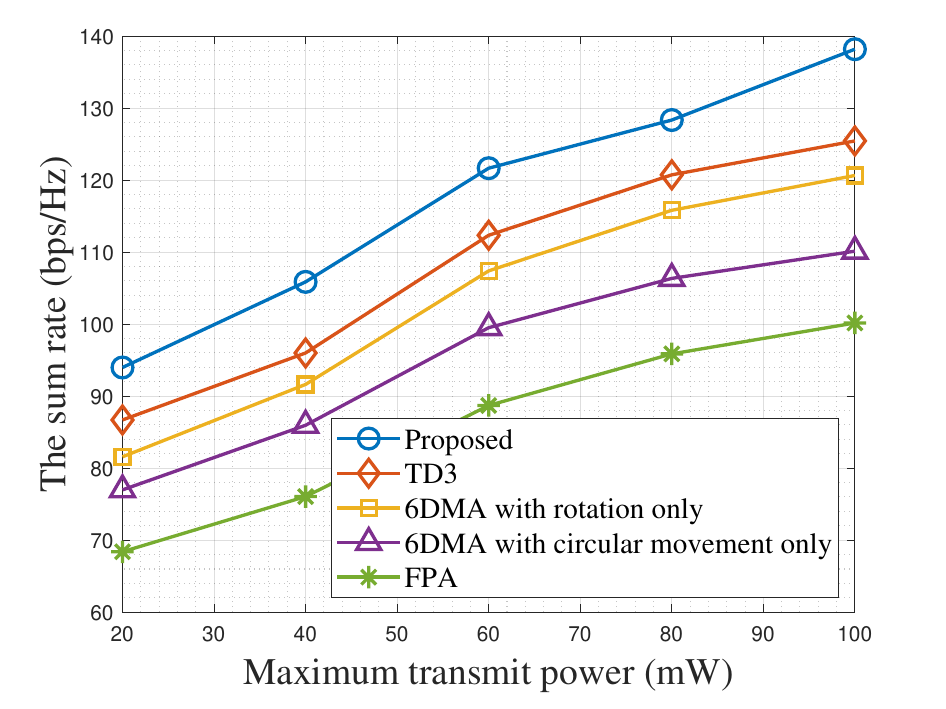}
    \caption{The sum rate versus the transmit power of the BS.}
    \label{transmitpower}
\end{figure}

\begin{figure}[htbp]
    \centering
    \includegraphics[width=0.8\linewidth]{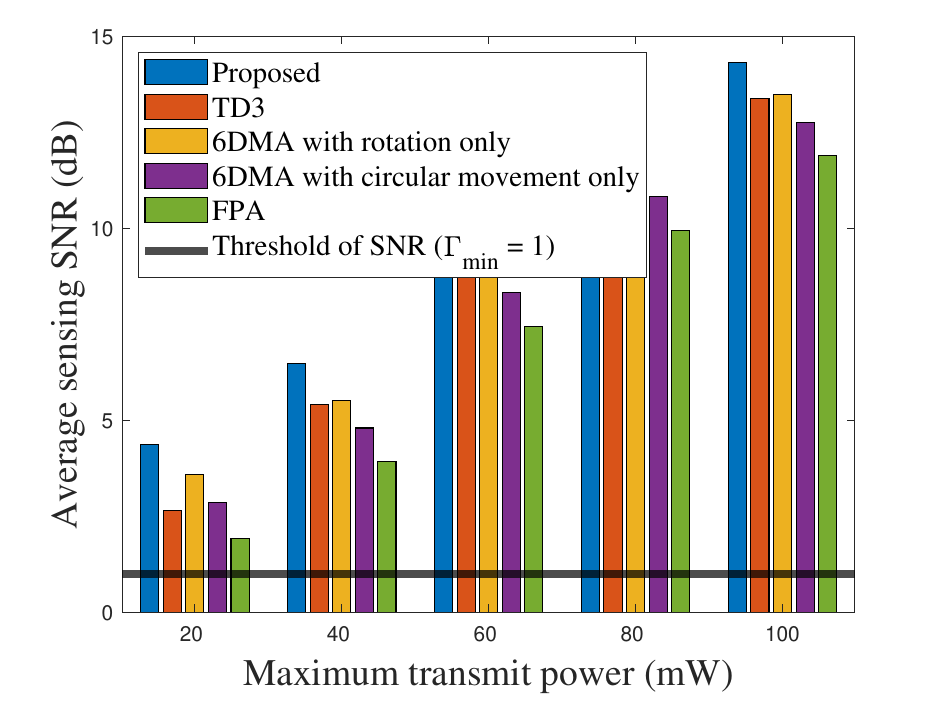}
    \caption{The average SNR versus the transmit power of the BS.}
    \label{AveSnr}
\end{figure}

The relationship between the maximum transmission power of the BS and the total data transmission rate is shown in Fig.~\ref{transmitpower}. We can observe that due to the rotation of the joint optimization 6DMA and the effectiveness of the location, the data rate of the proposed scheme is always better than the other comparison schemes. The reason is that under this scheme, the antenna surface has a higher spatial degree of freedom, which is better enough to adjust the beam direction of the transmitted signal and enhance the signal strength.
Fig.~\ref{AveSnr} shows the average sensing SNR of the five schemes under different transmit powers. It can be seen from the Fig.~\ref{AveSnr} that the average SNR can meet the sensing requirements $\Gamma_{\min} = 1\ \text{dB}$. The proposed scheme, through dynamically adjusting the 6DMA surface position and rotation, still has a relatively high sensing intensity even at a lower transmit power, indicating that it has stronger robustness in sensing and recognition under harsh channel conditions.

\begin{figure}
    \centering
    \includegraphics[width=0.8\linewidth]{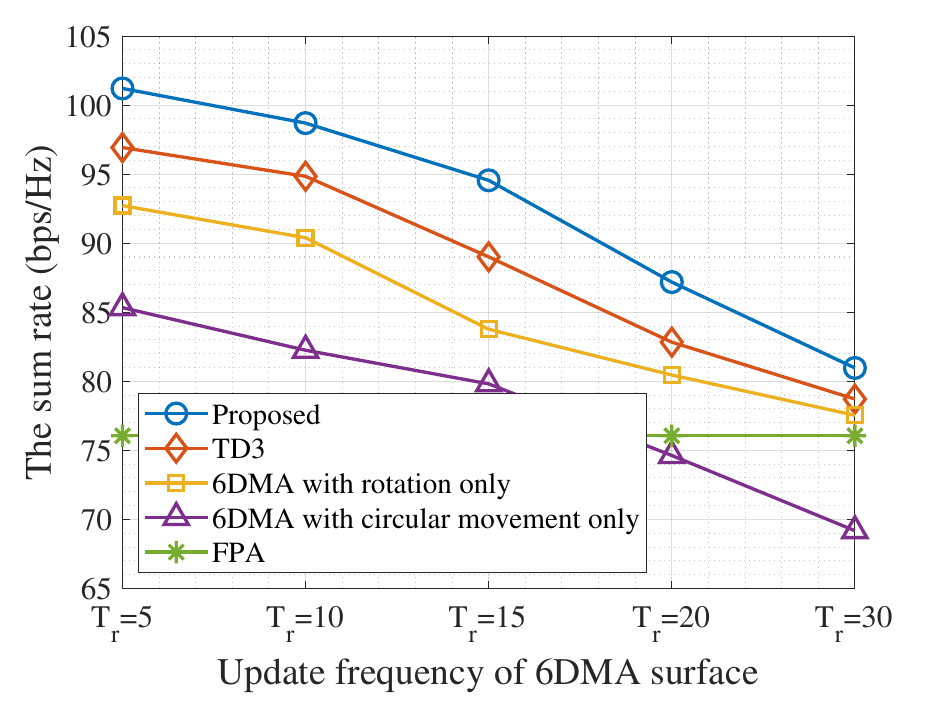}
    \caption{The sum rate verus update frequency of 6DMA.}
    \label{update_frequency}
\end{figure}

Fig.~\ref{update_frequency} shows the relationship between the sum rate of different schemes and the update frequency of 6DMA. In this paper, the total number of task time slots is 60, so we set $T_r \in \{ 5,10,15,20,30\}$. We can observe that under different $T_r$ values, the sum rate of the proposed scheme is higher than that of other schemes. However, as $T_r$ increases, the achievable total rate of each scheme gradually decreases except for FPA, and the magnitude of the decrease generally gradually increases. This is because the low update frequency of 6DMA makes it unable to adjust its position and rotation in a timely manner to adapt to the new system state.

\begin{table}[htbp]
	\centering
    \renewcommand\arraystretch{1.25}
	\caption{{Inference latency of each agent.}}
    \label{inference_latency}
    \resizebox{\linewidth}{!}{
	\begin{tabular}{|c|c|c|c|c|}
		\hline
		\multicolumn{2}{|c|}{\textbf{Agent}} & {\makecell[c]{\textbf{Average} \\ \textbf{latency (ms)}}} & {\makecell[c]{\textbf{Max} \\ \textbf{latency (ms)}}} & {\makecell[c]{\textbf{P99} \\ \textbf{latency (ms)}}}  \\ \hline
		\multirow{4}{*}{UAV Agent} & 1 & 0.27 & 0.771 & 0.514  \\ \cline{2-5}
		& 2 & 0.207 & 0.744 & 0.443  \\ \cline{2-5}
		& 3 & 0.188 & 0.613 & 0.365  \\ \cline{2-5}
		& 4 & 0.185 & 0.701 & 0.387  \\ \hline
		\multicolumn{2}{|c|}{Beamforming Agent} & 0.185 & 0.741 & 0.396  \\ \hline
		\multicolumn{2}{|c|}{6DMA Agent} & 0.288 & 0.629 & 0.535  \\ \hline
	\end{tabular}
    }
\end{table}
Table~\ref{inference_latency} shows the inference latencies per time slot of the proposed scheme on the experimental platform. As can be seen from the table, the average latency, the max latency and the P99 latency are less than 1 ms, where P99 latency refers to the response time such that $99\%$ of all requests have a response time less than or equal to this value. Taking a 100Hz control loop system as an example, the inference duration does not exceed $10\%$ of the control cycle, leaving sufficient margin for sensor data acquisition and communication, thus the proposed scheme is suitable for highly dynamic ISAC networks.

\section{Conclusion} \label{sec:conclusion}
In this paper, we propose a BS with 6DMA assisted ISAC for LAE. A HDRL algorithm based on MATD3 is used to jointly optimize the rotation and location of 6DMA, the trajectory of UAVs and the beamforming matrix of BS transmitting signals. The Agent of the first layer changes the rotation and location of 6DMA every $T_r$ time slots, and the agents of the second layer optimize the flight direction of UAVs and the beamforming matrix of BS transmitting signals every time slot. The numerical results verify that the proposed algorithm is significantly better than the algorithm based on TD3, the FPA scheme, and the partially movable schemes in the total data transmission rate and average SNR of the system, which proves the effectiveness of the proposed algorithm.
Multi-UAV flying will become common in the future LAE, and the number of UAVs varies over time. Therefore, it is valuable to conduct future research on multi-agent algorithms that can adapt to variable numbers of UAVs in LAE.

\bibliographystyle{IEEEtran}
\bibliography{ref}
\end{document}